\documentclass[aps,prl,twocolumn,superscriptaddress,floatfix]{revtex4-1}
\usepackage{makecell}
\usepackage{graphicx}
\usepackage{dcolumn}
\usepackage{bm}
\usepackage{times}
\usepackage{color}
\usepackage{appendix}
\usepackage[colorlinks=true,linkcolor=blue,anchorcolor=blue,citecolor=blue,urlcolor=blue ]{hyperref}
\usepackage{gensymb}
\usepackage{amsmath}
\usepackage{threeparttable}
\usepackage{ulem}

\begin{document}
\bibliographystyle{apsrev4-1}
\title{Critical behavior in the Mn$_{5}$Ge$_{3}$ ferromagnet }

\author{Jun-Fa Lin}
\affiliation{Department of Physics, Renmin University of China, Beijing 100872, P. R. China}
\affiliation{Beijing Key Laboratory of Opto-electronic Functional Materials $\&$ Micro-nano Devices, Renmin University of China, Beijing 100872, P. R. China}

	\author{Huan Wang}
\affiliation{Department of Physics, Renmin University of China, Beijing 100872, P. R. China}
\affiliation{Beijing Key Laboratory of Opto-electronic Functional Materials $\&$ Micro-nano Devices, Renmin University of China, Beijing 100872, P. R. China}

	\author{Sheng Xu}
\affiliation{Department of Physics, Renmin University of China, Beijing 100872, P. R. China}
\affiliation{Beijing Key Laboratory of Opto-electronic Functional Materials $\&$ Micro-nano Devices, Renmin University of China, Beijing 100872, P. R. China}

	\author{Xiao-Yan Wang}
\affiliation{Department of Physics, Renmin University of China, Beijing 100872, P. R. China}
\affiliation{Beijing Key Laboratory of Opto-electronic Functional Materials $\&$ Micro-nano Devices, Renmin University of China, Beijing 100872, P. R. China}

	\author{Xiang-Yu Zeng}
\affiliation{Department of Physics, Renmin University of China, Beijing 100872, P. R. China}
\affiliation{Beijing Key Laboratory of Opto-electronic Functional Materials $\&$ Micro-nano Devices, Renmin University of China, Beijing 100872, P. R. China}

	\author{Zheng-Yi Dai}
\affiliation{Department of Physics, Renmin University of China, Beijing 100872, P. R. China}
\affiliation{Beijing Key Laboratory of Opto-electronic Functional Materials $\&$ Micro-nano Devices, Renmin University of China, Beijing 100872, P. R. China}

	\author{Jing Gong}
\affiliation{Department of Physics, Renmin University of China, Beijing 100872, P. R. China}
\affiliation{Beijing Key Laboratory of Opto-electronic Functional Materials $\&$ Micro-nano Devices, Renmin University of China, Beijing 100872, P. R. China}

	\author{Kun Han}
\affiliation{Department of Physics, Renmin University of China, Beijing 100872, P. R. China}
\affiliation{Beijing Key Laboratory of Opto-electronic Functional Materials $\&$ Micro-nano Devices, Renmin University of China, Beijing 100872, P. R. China}

	\author{Yi-Ting Wang}
\affiliation{Department of Physics, Renmin University of China, Beijing 100872, P. R. China}
\affiliation{Beijing Key Laboratory of Opto-electronic Functional Materials $\&$ Micro-nano Devices, Renmin University of China, Beijing 100872, P. R. China}

	\author{Xiao-Ping Ma}
\affiliation{Department of Physics, Renmin University of China, Beijing 100872, P. R. China}
\affiliation{Beijing Key Laboratory of Opto-electronic Functional Materials $\&$ Micro-nano Devices, Renmin University of China, Beijing 100872, P. R. China}

\author{Tian-Long Xia}\email{tlxia@ruc.edu.cn}
\affiliation{Department of Physics, Renmin University of China, Beijing 100872, P. R. China}
\affiliation{Beijing Key Laboratory of Opto-electronic Functional Materials $\&$ Micro-nano Devices, Renmin University of China, Beijing 100872, P. R. China}

\date{\today}

\begin{abstract}
High-Curie-temperature ferromagnets are promising candidates for designing new spintronic devices. Here we have successfully synthesized a single-crystal sample of the itinerant ferromagnet  Mn$ _{5}$Ge$_{3}$ used flux method and its critical properties were investigated by means of bulk dc-magnetization at the boundary between the ferromagnetic (FM) and paramagnetic (PM)  phase. Critical exponents $ \beta=0.336 \pm 0.001 $  with a critical temperature $ T_{c}=300.29  \pm  0.01 $ K and $ \gamma=1.193 \pm 0.003  $ with $ T_{c} = 300.15 \pm 0.05 $ K are obtained by the modified Arrott plot, whereas $ \delta = 4.61 \pm 0.03  $ is deduced by a critical isotherm analysis at $ T_{c} = 300 $ K. The self-consistency and reliability of these critical exponents are verified by the Widom scaling law and the scaling equations. Further analysis reveals that the spin coupling in Mn$ _{5}$Ge$_{3}$ exhibits three-dimensional Ising-like behavior. The magnetic exchange is found to decay as $ J(r)\approx r^{-4.855} $ and the spin interactions are extended beyond the nearest neighbors, which may be related to different set of Mn--Mn interactions with unequal magnitude of exchange strengths. Additionally, the existence of noncollinear spin configurations in Mn$ _{5} $Ge$ _{3} $ results in a small deviation of obtained critical exponents from those for standard 3D-Ising model.

\end{abstract}

\maketitle

\section{Introduction}
Itinerant ferromagnets have been extensively studied due to their exotic physical properties, for example, unconventional superconductivity, non-Fermi-liquid behavior, exotic magnetic states, or quantum critical behavior \citep{Shimizu_1981,10.1038/35020500, 10.1038/35098048, 10.1038/nature05056, 10.1038/35106527,10.1038/35007030,PhysRevB.55.8330}. The transition of paramagnetic (PM) to ferromagnetic (FM) phase with decreasing temperature is regarded as a canonical example of a second order phase transition. Specifically, many experimental and theoretical studies have looked at enigmatic phenomena related to a quantum phase transition (QPT) between FM and PM states. Furthermore, a second order (critical) QPT in itinerant-electron systems are believed to be responsible for enigmatic quantum phases like magnetically mediated superconductivity and non-Fermi liquid behavior in materials such as URhAl \citep{PhysRevB.97.064423}, MnSi \citep{PhysRevB.55.8330,PhysRevB.91.024403}, and UGe$ _{2} $ \citep{PhysRevLett.89.147005}. Here we focus on a classical critical behavior of the magnetization around  a ferromagnetic transition temperature from which the type of the magnetic phase transition and the nature of spin-spin interactions can be studied\citep{PhysRevB.96.144429,10.1038/srep22397}.

In Mn--based materials, Mn$ _{5} $Ge$ _{3} $ is prominent ferromagnet with the magnetic ordering at $T _{c}=297 $ K, close to the room temperature \citep{10.1143/JPSJ.18.773}. Due to its considerable spin-polarization\citep{10.1002/pssb.200510030} and the possibility to epitaxially grow on semiconductors, Mn$ _{5} $Ge$ _{3} $ has been subjected to wide investigation as a potential candidate for efficient spin injector in spintronics\citep{10.1063/1.4817372,PhysRevB.70.235205}. Furthermore, the fundamental properties of Mn$ _{5} $Ge$ _{3} $, especially its magnetic structure, have been investigated in detail\citep{Jackson_1965,10.1143/JPSJ.18.773,10.1002/pssa.2210340233,Forsyth_1990}. Mn$ _{5} $Ge$ _{3} $ holds $ D8_{8} $ structure same as Mn$ _{5} $Si$ _{3} $ with the space group P6$ _{3} $/mcm, and the unit cell contains 6 Ge and 10 Mn atoms. Mn atom occupies two different Wyckoff positions: 4d (Mn\uppercase\expandafter{\romannumeral1}) and 6g (Mn\uppercase\expandafter{\romannumeral2}) with the magnetic moments of 1.96 $ \mu_{B} $ and 3.23 $ \mu_{B} $, respectively\citep{Forsyth_1990}. The magnetic moment direction of Mn\uppercase\expandafter{\romannumeral1} and Mn\uppercase\expandafter{\romannumeral2} atoms has been reported to be parallel to the c axis of the hexagonal structure from $T_{c}$ to around 70 K\citep{PhysRevB.104.064416}. The difference among various Mn--Mn interactions leads to an anisotropic exchange and complex magnetic ordering in different temperature regimes.

The theoretical calculations have demonstrated that Mn$ _{5} $Ge$ _{3} $ has two competing phases with collinear and noncollinear spin configurations \citep{10.1002/pssa.200673014}. The transition temperature from collinear to noncollinear magnetism in Mn$ _{5} $Ge$ _{3} $ is about 70 K \citep{10.1063/1.1633684,10.1021/nl303645k}, quite similiar as the situation in Mn$ _{5} $Si$ _{3} $ where its magnetic transition from a  ``high-temperature'' collinear AFM spin state to a low-temperature noncollinear antiferromagnetic (AFM) spin state  occurs at almost the same temperature (66 K)\citep{Brown_1992,Brown_1995,10.1038/ncomms4400}. Furthermore, the density-functional theory (DFT) calculations reveal that in general the interaction between the nearest neighbors Mn\uppercase\expandafter{\romannumeral1}--Mn\uppercase\expandafter{\romannumeral1} is ferromagnetic and it is much stronger than the interaction between the Mn\uppercase\expandafter{\romannumeral1}--Mn{\uppercase\expandafter{\romannumeral2} or Mn\uppercase\expandafter{\romannumeral2}--Mn\uppercase\expandafter{\romannumeral2} atoms. Neighboring Mn\uppercase\expandafter{\romannumeral2} atoms are ferromagnetically coupled in the fully relaxed unit cell. Nevertheless, by applying compressive strain, the Mn\uppercase\expandafter{\romannumeral2}--Mn\uppercase\expandafter{\romannumeral2} atomic distance decreases and the corresponding FM coupling is suppressed, then transformed into AFM coupling \citep{PhysRevB.104.064416, 10.1063/1.3134482}. In addition, it is believed that the coexistence of AFM and FM coupling observed at low temperature may lead to the low-temperature noncollinear spin configuration which is commonly observed in Mn$ _{5} $Ge$ _{3} $ thin film or nanowire \citep{PhysRevB.104.064416,10.1063/1.1633684,10.1021/nl303645k,10.1063/1.3134482}.

Mn$ _{5} $Ge$ _{3} $ has been widely studied for its magnetocaloric properties due to the large effect near room temperature, which makes it a potential candidate for magnetic refrigeration with the advantage of being environment friendly  \citep{10.1002/pssb.201248474,TOLINSKI20141,KANG2017931,LALITA2021159908}. Though the magnetothermal properties of Mn$_{5}  $Ge$ _{3} $ have been extensively investigated, further studies are desired to understand the intrinsic magnetic interactions, especially the critical behavior in Mn$ _{5} $Ge$ _{3} $ single crystal. In this paper, we investigate the critical behavior of flux-grown Mn$ _{5} $Ge$ _{3} $ crystal with various techniques, such as modified Arrott plot, Kouvel--Fisher plot, and critical isotherm analysis. Our analyses indicate that the obtained critical exponents [ $ \beta=0.336 \pm 0.001 $  ( $T_{c}=300.29  \pm  0.01 $ K), $ \gamma=1.193 \pm 0.003 $ ($ T_{c} = 300.15 \pm 0.05$ K), and $ \delta = 4.61 \pm 0.03$  ($  T_{c}=300 $ K)] are close to those values obtained from the renormalization group calculation for the three-dimensional Ising model\citep{KAUL19855}. The spin interaction decays with distance $ r $ as $ J(r) \approx r^{-(3+\sigma)} (\sigma = 1.855) $, indicating that the spin interaction is extended beyond the nearest neighbors (\textit{NN})\citep{PhysRevLett.29.917}, which may result from the different Mn--Mn interactions with unequal magnitude of exchange strengths. Subsequently, the existence of noncollinear spin configurations in Mn$ _{5} $Ge$ _{3} $ leads to a small deviation of obtained critical exponents from those for standard 3D-Ising model.

\section{experimental details}
 High-quality Mn$ _{5} $Ge$ _{3} $ single crystals were grown by the flux method starting from a mixture of pure elements Mn powder, Ge powder, and In granular with a molar ratio of $ 5:3:200 $. The starting materials were loaded into corundum crucible which was sealed in an evacuated quartz tube. Then the tube was put into furnace and heated to 950 $ \celsius $ over 10 h, held for 20 h, then slowly cooled down to 350 $ \celsius $ at a rate of 2 $ \celsius$ / h. Finally, crystals with metallic luster were obtained by centrifugation to remove excess flux at 350 $ \celsius $. Elemental analysis was performed using energy dispersive X-ray spectroscopy (EDX, Oxford X-Max 50). The determined atomic proportion was consistent with the composition of Mn$ _{5} $Ge$ _{3} $ within instrumental error. The single crystal and powder X-ray diffraction (XRD) patterns were carried out by a Bruker D8 Advance X-ray diffractometer using Cu K$ _{\alpha} $ radiation. TOPAS-4.2 was employed for the refinement. The measurements of resistivity and magnetic properties were performed on the Quantum Design physical property measurement system (QD PPMS-14T) and magnetic property measurement system (MPMS-3). The M(H) curves are measured at interval $ \bigtriangleup T $ = 1 K, and $ \bigtriangleup T $ = 0.5 K when approaching $ T_{c} $. The applied magnetic field $ H_{a} $ has been corrected into the internal field as $ H = H_{a} - NM $, where $ M $ is the measured magnetization and $ N $ is the demagnetization factor. The corrected $ H $ was used for the analysis of critical behavior.

\section{Results and discussions}
\begin{figure}[htbp]
	\centering
	\includegraphics[width=0.51\textwidth]{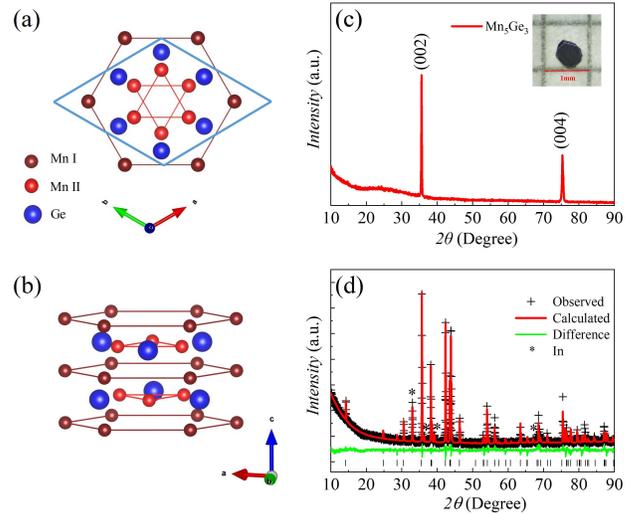}
	\caption{ Crystal structure of Mn$_5$Ge$ _{3} $ with the space group P6$ _{3} $/mcm (No.193) from (a) top and (b) side views. The unit cell is enclosed by blue solid lines. Nonequivalent Mn sites are labeled as Mn\uppercase\expandafter{\romannumeral1} and Mn\uppercase\expandafter{\romannumeral2}, respectively. (c) Single crystal XRD pattern of Mn$ _{5} $Ge$ _{3} $ at room temperature. Inset: the picture of Mn$ _{5} $Ge$ _{3} $ crystal. (d) Powder XRD pattern with refinement. The vertical tick marks represent Bragg reflections of the space group P6$ _{3} $/mcm.}
	\label{Fig1} 	
\end{figure}
The intermetallic compound Mn$ _{5} $Ge$ _{3} $ is ferromagnetic below its Curie temperature, and its crystal structure is of the $ D8_{8} $ type as shown in Figs. \ref{Fig1}(a) and (b), with lattice parameters $ a=b=7.184(2) $ \AA  $  $ and $ c=5.053(2) $ \AA \,at ambient temperature. The local environment of the Mn atoms in Mn$ _{5} $Ge$ _{3} $ is illustrated in Figs.\ref{Fig1} (a) and (b): Mn\uppercase\expandafter{\romannumeral1} has two \textit{NN} Mn\uppercase\expandafter{\romannumeral1} (at 2.522 \AA ) and six next nearest neighbours (\textit{NNN}) Mn\uppercase\expandafter{\romannumeral2} (at 3.059 \AA), whereas Mn\uppercase\expandafter{\romannumeral2} has two \textit{NN} Mn\uppercase\expandafter{\romannumeral2} (at 2.976 \AA), four \textit{NNN} Mn\uppercase\expandafter{\romannumeral2} (at 3.051 \AA), and four \textit{NNN} Mn\uppercase\expandafter{\romannumeral1} (at 3.059 \AA)\citep{Forsyth_1990}. Fig. \ref{Fig1}(c) presents the single-crystal XRD pattern, where only ($ 00l $) peaks are detected, indicating the crystal surface is normal to c-axis and the hexagon-shaped surface is parallel to the $ ab $ plane. The inset shows the picture of as grown Mn$ _{5} $Ge$ _{3} $ crystal after handling with dilute HCl acid (7 wt \%) to remove the residual flux on the sample surface. Furthermore, the powder XRD pattern of Mn$ _{5} $Ge$ _{3} $ is shown in Fig. \ref{Fig1}(d), in which the observed peaks are well fitted with the P6$ _{3} $/mcm space group. Indium flux is also observed on the pattern since the XRD experiments are carried out on the powders from crushing small single crystals.
\begin{figure}[htbp]
	\centering
	\includegraphics[width=0.48\textwidth]{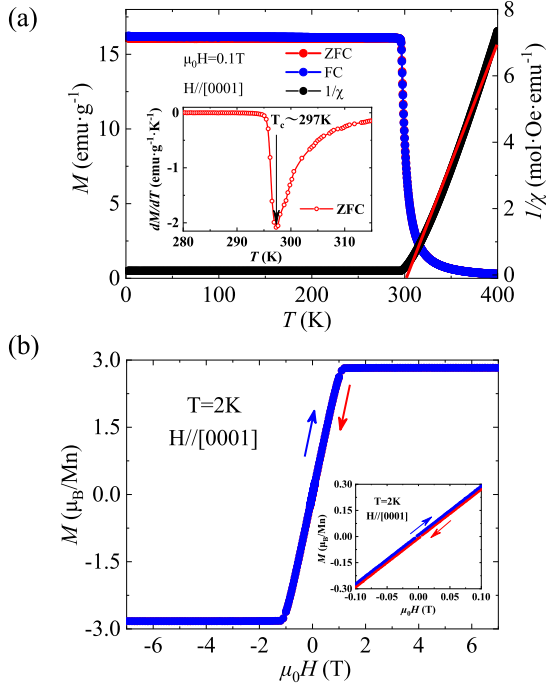}
	\caption{ (a) The temperature dependence of magnetization (blue and red dotted line) and inverse susceptibility (black dotted line) for Mn$_5$Ge$ _{3} $ measured under the magnetic field $ H=0.1 $ T applied along the $ c $ axis with zero-field-cooling (ZFC) and field-cooling (FC) modes. The red solid line is fitted by the modified Curie-Weiss law. Inset shows the first-order derivative of magnetization ($ dM/dT $) vs. $ T $. (b) The isothermal magnetization at 2 K. Inset shows the enlarged part at low field.   }
	\label{Fig2} 	
\end{figure}

The critical temperature $ T_{c} $ can be roughly determined by the temperature dependence of magnetization $ M(T) $. Fig. \ref{Fig2}(a) shows the $ M(T) $ curves for Mn$ _{5} $Ge$ _{3} $ under zero-field-cooling (ZFC) and field-cooling (FC) with an applied field $ H=0.1 $ T, which is parallel to the $ c $ axis. The $ M(T) $ curve exhibits an abrupt decline with the increase of temperature, corresponding to the FM-PM transition. The inset of Fig. \ref{Fig2}(a) presents $ dM/dT $ vs. $ T $, where $ T_{c}=297 $ K is determined from the minimum of the $ dM/dT $ curve. The corresponding $ T $-dependence of the inverse susceptibility $ 1/\chi $ is also shown in Fig. \ref{Fig2}(a). Excellent fitting to the data in the high-temperature range was obtained using the modified Curie-Weiss law 
\begin{equation}\label{equ1}
\centering
\chi=\chi_{0}+\frac{C}{T-\theta}
\end{equation} 
where $ \chi_{0} $ is the temperature-independent susceptibility, $ C $ is the Curie-Weiss constant, and $ \theta $ is the Weiss temperature. The values of the parameters obtained from the fitting are $ C=18.482 $ emu$ \cdot $ K/mol$ \cdot $Oe, $ \theta = 301.41 $ K, and $ \chi_{0} =0.040 $ emu/mol$ \cdot $Oe with $ H//c $. The positive $ \theta $ confirms the FM interaction among Mn atoms. Also, effective magnetic moment per Mn atom of Mn$ _{5} $Ge$ _{3} $ is $5.44 \mu_{B}$, which can be calculated from Curie's constant through, $ \mu_{eff} =\sqrt{8C/N} $, with $ N $ being number of moment bearing ions per formula unit (here $ N=5 $). Fig. \ref{Fig2}(b) displays the isothermal magnetization measured at $ T=2 $ K, and the saturation moment $ M_{s} $ estimated by linear extrapolation of high field magnetic isotherm is $ \sim2.82 \mu_{B} $/Mn, which is consistent with previously reported value\citep{PhysRevLett.96.037204}. The inset in Fig. \ref{Fig2}(b) shows the $ M(H) $ in the low-field region and the little hysteresis with the coercive forces $ H_{c} = 25  $ Oe has been observed. On the other hand, the number of magnetic carriers ($ P_{c} $) deduced from the effective Bohr magneton number ($P_{eff}$, effective moment $\mu_{eff}$=$P_{eff}$$\cdot$$\mu_{B}$) using, $ P_{eff}^{2} = P_{c}(P_{c}+2) $ with $ P_{c}/2=S $ (the effective spin per atom), is 4.53. The Rhodes-Wohlfarth ratio (RWR) defined as $ P_{c}/P_{s} $, where $ P_{s} $ is also the number of magnetic carriers per atom deduced from direct low temperature measurements of the saturation magnetization, is usually estimated to figure out whether the magnetism is localized or itinerant in origin. The RWR equals 1 for a localized system and larger in an itinerant system \citep{10.1098/rspa.1963.0086,WOHLFARTH1978113,MORIYA19791}. In Mn$ _{5} $Ge$ _{3} $, the RWR equals 1.61 indicating a possible weak itinerant character. 

The temperature dependence of resistivity $ \rho $ for the Mn$_{5}$Ge$ _{3} $ single crystal is shown in Fig. \ref{Fig3}, and the resistivity monotonically decreases from about 245 $ \mu\Omega\cdot  $cm at 320 K down to about 2.9 $ \mu\Omega\cdot  $cm at 2 K, confirming the metallic characteristic of Mn$_{5}$Ge$ _{3} $. The residual resistivity ratio (RRR), $ \rho $ (300 K) / $ \rho $ (2 K) $ \sim $ 83, indicates high quality of the sample. There is a clear kink at 297 K in the $ \rho$  $ \textendash$ T curve, which is mainly ascribed to the transition from paramagnetic to ferromagnetic state, consistent with the result of the $ M(T) $ curve.

In addition to the magnetic phase transition at Curie temperature, there is a cusp near $ T_{Mn} \sim 70 $ K, which is clearly shown in the d$ \rho $ / dT curve. Such a cusp is commonly observed in Mn$ _{5} $Ge$ _{3} $ thin films and nanostructures but never observed in bulk crystal before\citep{10.1063/1.1633684,10.1021/nl303645k,PhysRevB.104.064416}. For the first time, we observed it in the single crystals. The appearance of the cusp is usually considered as the emergence of another magnetic ordering. The spatial distribution of magnetization density in Mn$ _{5} $Ge$ _{3} $ has been determined from a polarized neutron diffraction study on single crystals, which revealed two crystallographically nonequivalent Mn sublattices (Mn\uppercase\expandafter{\romannumeral1} and Mn\uppercase\expandafter{\romannumeral2})\citep{Forsyth_1990}. The DFT calculations also suggest the coexistence of two competing ferromagnetic states, collinear and noncollinear spin configurations, in the Mn$ _{5} $Ge$ _{3} $ lattice\citep{10.1002/pssa.200673014}. Here, the observed cusp in the d$ \rho $ / dT curve is attributed to a possible magnetic transition between collinear and noncollinear ferromagnetic states. A similar scenario occurs in the antiferromagnetic Mn$ _{5} $Si$ _{3} $, where a magnetic transition from collinear to noncollinear antiferromagnetic spin states occurrs at $ T \sim 66 $ K, almost the same as in Mn$ _{_{5}} $Ge$ _{3} $. For Mn$ _{5} $Si$ _{3} $, the noncollinear magnetic structure of the AF1 phase is stable below 66 K, where the noncollinearity is attributed to topological frustration, and the collinear AF2 phase between 66 K and 99 K show the collinear AFM spin configurations\citep{Brown_1992,Brown_1995,Leciejewicz2008,10.1038/ncomms4400}.
\begin{figure}[htbp]
	\centering
	\includegraphics[width=0.48\textwidth]{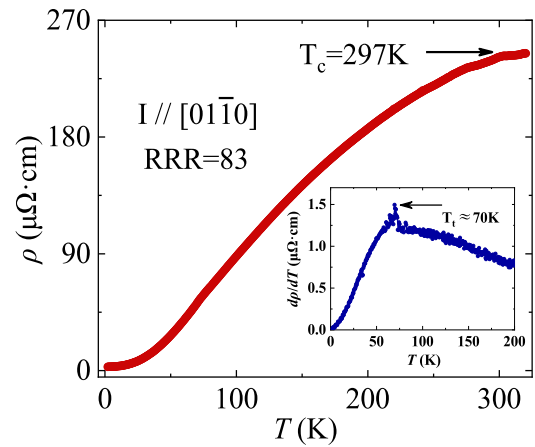}
	\caption{Temperature dependence of the resistivity for the Mn$ _{5} $Ge$ _{3} $ single crystal, and the corresponding first-order derivative $ d\rho/dT $ is shown on the inset.}
	\label{Fig3} 	
\end{figure}

\begin{figure}[htbp]
	\centering
	\includegraphics[width=0.45\textwidth]{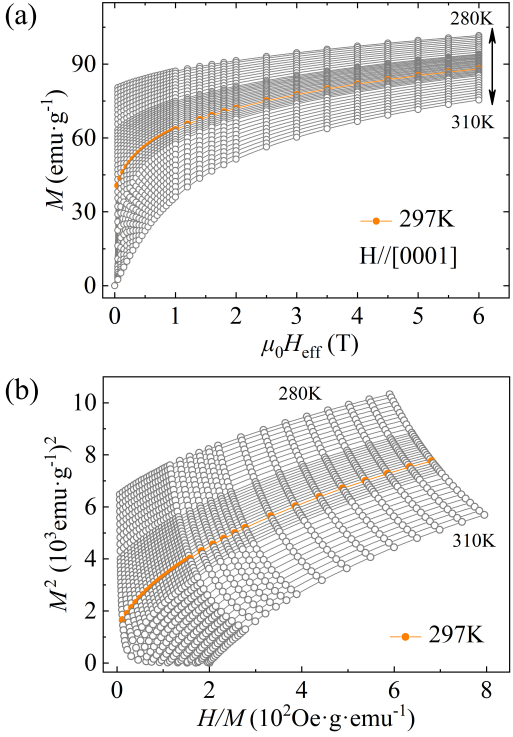}
	\caption{ (a) The initial isothermal magnetization curves  around $ T_{c} = 297 $ K (in an orange symbol and line) for Mn$ _{5} $Ge$ _{3} $. (b) Arrott plots of $ M^{2} $ vs. $ H/M $ around $ T_{c} $ for Mn$_{5}$Ge$ _{3} $ (the $ M(H) $ curves are measured at interval $ \Delta T = 1 $ K, and $ \Delta T = 0.5 $ K when approaching $ T_{c} $).   }
	\label{Fig4} 	
\end{figure}

It is well known that the critical behavior for a second-order phase transition can be characterized by a series of interrelated critical exponents. In the vicinity of the critical point, the divergence of correlation length $ \xi = \xi_{0}|(T-T_{c})/T_{c}|^{-\nu} $ leads to universal scaling laws for the spontaneous magnetization $ M_{s} $ and initial susceptibility $ \chi_{0} $. The spontaneous magnetization $ M_{s} $ below $ T_{c} $, the inverse initial susceptibility $ \chi_{0}^{-1} $ above $ T_{c} $, and the measured magnetization $ M(H) $ at $ T_{c} $ are characterized by a series of critical exponents $ \beta$, $\gamma $, and $ \delta $, respectively. Subsequently, the mathematical definitions of the exponents from magnetization are described as\citep{Fisher_1967,stanley1971phase}: 
\begin{equation}\label{equ2}
\centering
M_{s}(T) = M_{0}(-\varepsilon)^{\beta},    \varepsilon < 0, T < T_{c}
\end{equation} 
\begin{equation}\label{equ3}
\centering
\chi_{0}^{-1} = (h_{0}/M_{0}) \varepsilon^{\gamma}, \varepsilon > 0, T > T_{c}
\end{equation} 
\begin{equation}\label{equ4}
\centering
M = DH^{1/\delta}, \varepsilon = 0, T = T_{c}
\end{equation}
where $ \varepsilon = (T - T_{c})/T_{c} $ is the reduced temperature, and $ M_{0} $, $ h_{0}/m_{0} $, and $ D $ are the critical amplitudes. Generally, in the asymptotic critical region ($ |\varepsilon| < 0.1 $), these critical exponents should follow the Arrott--Noakes equation of state \citep{PhysRevLett.19.786}:
\begin{equation}\label{equ5}
\centering
(H/M)^{1/\gamma} = (T-T_{c})/T_{c} + (M/M_{1})^{1/\beta}
\end{equation}
where $ M_{1} $ is a constant.

According to the fitting of $ M_{s}(T) $ and $ \chi_{0}^{-1}(T) $ curves by the modified Arrott plot of $ M^{1/\beta} $ vs. $ (H/M)^{1/\gamma} $, the critical exponents $ \beta $ and $ \gamma $ are obtained. Moreover, $ \delta $ is yielded directly by the $ M(H) $ at the critical temperature $ T_{c}$ based on the Eq.(\ref{equ4}).

In order to clarify the nature of the PM--FM transition in Mn$ _{5} $Ge$ _{3} $, we measured the isothermal $ M(H) $ in the temperature range from $ T=280 $ K to $ T=310 $ K, as shown in Fig. \ref{Fig4}(a). Universally, the critical exponents can be determined by the Arrott plot \citep{PhysRev.108.1394}. For the Landau mean-field model with $ \beta=0.5 $ and $ \gamma=1.0 $ \citep{KAUL19855}, the Arrott--Noakes equation of state evolves into Arrott equation $ H/M = A + BM^{2} $. The Arrott plot of $ M^{2} $ vs. $ H/M $ for Mn$_{5}$Ge$ _{3} $ is depicted in Fig. \ref{Fig4}(b). According to Banerjee's criterion \citep{BANERJEE196416}, the slope of line in the Arrott plot can estimate the order of magnetic transition: the negative slope corresponds to the first-order transition while the positive corresponds to the second order. Therefore, the concave downward curvature clearly indicates that the PM--FM transition in Mn$ _{5} $Ge$ _{3} $ is the second-order phase transition, in agreement with the specific heat measurement \citep{TOLINSKI20141}. According to the Arrott plot, isotherms plotted in the form of $ M^{2} $ vs. $ H/M $ constitute a series of parallel straight lines around $ T_{c} $, and the isotherm at the critical temperature $ T_{c} $ should pass through the origin \citep{PhysRev.108.1394}. At the same time, $ \chi_{0}^{-1}(T) $ and $ M_{s}(T) $ can be directly obtained from the intercepts on the $ H/M $ axis and positive $ M^{2} $ axis, respectively. All the $ M^{2} $ vs. $ H/M $ curves in Fig. \ref{Fig4}(b) show quasi-straight line with positive slops in high field range. However, all lines show a downward curvature and are not parallel to each other, indicating that the framework of the Landau mean-field model is not applicable in Mn$ _{5} $Ge$ _{3} $.
\begin{figure}[htbp]
	\centering
	\includegraphics[width=0.51\textwidth]{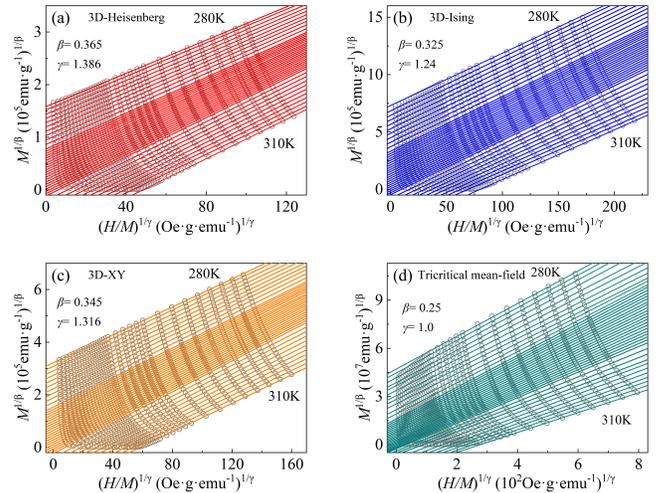}
	\caption{The isotherms of $ M^{1/\beta} $ vs. $ (H/M)^{1/\gamma} $ with parameters of (a) 3D-Heisenberg model, (b) 3D-Ising model, (c) 3D-XY model, and (d) tricritical mean-field model. The straight lines are the linear fit of isotherms at different temperatures. }
	\label{Fig5} 	
\end{figure}
\begin{figure*}[htbp]
	\centering
	\includegraphics[width=\textwidth]{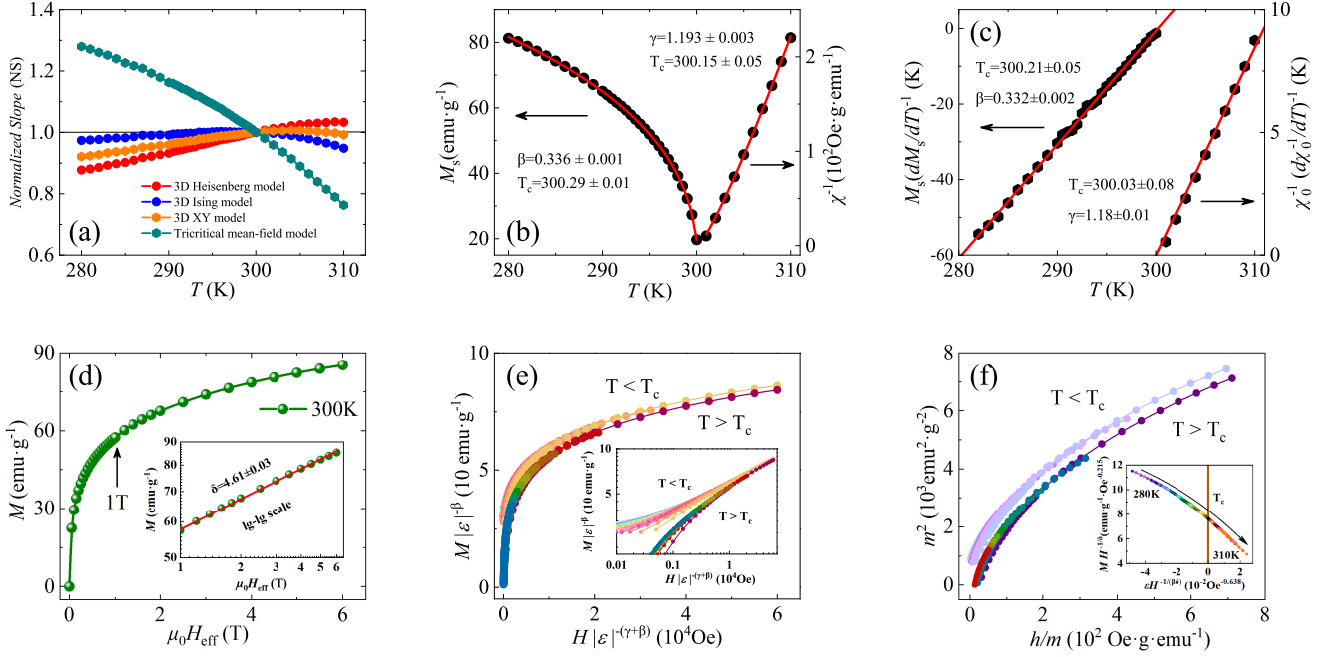}
	\caption{(a) The normalized slopes [$ NS = S(T)/S(T_{c}) $] as a function of temperature. (b) The temperature dependence of the spontaneous magnetization $ M_{s} $ (left) and the inverse initial susceptibility $ \chi_{0}^{-1} $ (right) for Mn$ _{5} $Ge$ _{3} $ with the fitting curves (red solid line). (c) Kouvel--Fisher plots of $M_{s}(dM_{s}/dT)^{-1}  $ (left) and $ \chi_{0}^{-1}(d\chi_{0}^{-1}/dT)^{-1} $ (right) with fitting curves for Mn$ _{5}$Ge$ _{3} $. (d) The isotherm $ M(H) $ plot collected at $ T_{c} = 300 $ K for Mn$ _{5} $Ge$ _{3} $. Inset: The same plot in lg-lg scale with the fitting curve. (e) Scaling plots of renormalized magnetization $ m  $ vs. renormalized field $ h $ below and above $ T_{c} $ for Mn$ _{5} $Ge$ _{3} $. Inset: the same plots in lg-lg scale. (f) The renormalized magnetization and field replotted in the form of $ m^{2} $ vs. $ h/m $ for Mn$ _{5} $Ge$ _{3} $. Inset: the rescaling of the $ M(H) $ curves by $ MH^{-1/\delta} $ vs. $ \varepsilon H^{-1/(\beta\delta)} $.   }
	\label{Fig6}
\end{figure*}

Thus, a modified Arrott plot of $ M^{1/\beta} $ vs. $ (H/M)^{1/\gamma} $ is necessary \citep{PhysRevLett.19.786}, and the modified Arrott plot is given by the Arrott--Noakes equation of state presented in Eq.\ref{equ5}. As shown in Figs.\ref{Fig5}(a--d), the modified Arrott plots are constructed based on four kinds of possible exponents in the 3D-Heisenberg model ($ \beta = 0.365, \gamma = 1.386 $), 3D-Ising model ($ \beta = 0.325, \gamma = 1.24 $), 3D-XY model ($ \beta = 0.345, \gamma = 1.316$), and tricritical mean-field model ($ \beta = 0.25, \gamma = 1.0 $) \citep{KAUL19855,10.1038/srep22397}. All these four constructions exhibit quasi-straight lines in high field region. Obviously, the unparallel lines in Fig. \ref{Fig5}(d)  indicate that the tricritical mean-field model is also not satisfied. However, all lines in the modified Arrott plots Figs.\ref{Fig5}(a--c) are almost parallel to each other. Actually, in an appropriate model, a series of parallel lines with the same slope in high field are obtained, where the slope $ S(T) = dM^{1/\beta}/d(H/M)^{1/\gamma}$. Furthermore, the normalized slope (NS) is defined as $ NS = S(T)/S(T_{c}) $, which enables us to identify the most suitable model by comparing the $ NS $ with the ideal value ``1''\citep{PhysRevB.81.144426}. Plot of $ NS $ vs. $ T $ for the four different models is shown in Fig. \ref{Fig6}(a). Obviously, the $ NS $ of 3D-XY model almost equals 1 above $ T_{c} $, while that of 3D-Ising model is near 1 below $ T_{c} $, indicating the critical behavior of Mn$ _{5} $Ge$ _{3} $ may not belong to a single universality class.

The iteration method is employed to extract the precise critical exponents $ \beta $ and $ \gamma $ \citep{PhysRevB.79.214426}. The linear extrapolation from the high field region to the intercepts with the axes $ M^{1/\beta} $ and $ (H/M)^{1/\gamma} $ yields reliable values of spontaneous magnetization $ M_{s}(T,0) $ and inverse initial susceptibility $ \chi_{0}^{-1}(T,0) $, which are plotted as a function of temperature. According to the fitting of data following the Eqs.(\ref{equ2}) and (\ref{equ3}), a set of $ \beta $ and $ \gamma $ are obtained, which are used to reconstruct a new modified Arrott plot. Hence, new $ M_{s}(T,0) $ and $ \chi_{0}^{-1}(T,0) $ are yielded from the linear extrapolation in the high field region. Then another set of $ \beta $ and $ \gamma $ can be obtained. This procedure is repeated until the values of $ \beta $ and $ \gamma $ do not change. Based on this method, the obtained critical exponents are independent on the initial parameters, which confirms these critical exponents are reliable and intrinsic. In this way, the final $ M_{s}(T) $ and $ \chi_{0}^{-1}(T) $ with solid fitting curve are presented in Fig. \ref{Fig6}(b). The critical exponents $ \beta = 0.336 \pm 0.001 $ with $ T_{c} = 300.29 \pm 0.01 $ K, and $ \gamma = 1.193 \pm 0.003 $ with $ T_{c} = 300.15 \pm 0.05 $ K for Mn$ _{5} $Ge$ _{3} $, are obtained. 

In order to check the accuracy of above analysis, the Kouvel--Fisher (KF) method is employed to fit the critical exponents and critical temperature, which is expressed as\citep{PhysRevB.79.214426,PhysRev.136.A1626}:

\begin{equation}\label{equ6}
\centering
\frac{M_{s}(T)}{dM_{s}(T)/dT} = \frac{T-T_{c}}{\beta}
\end{equation}
\begin{equation}\label{equ7}
\centering
\frac{\chi_{0}^{-1}(T)}{d\chi_{0}^{-1}(T)/dT} = \frac{T-T_{c}}{\gamma}
\end{equation}

The straight lines with slopes $ 1/\beta $ and $ 1/\gamma $ are obtained from $ M_{s}(T)/(dM_{s}(T)/dT) $ vs. $ T $ and $ \chi_{0}^{-1}(T)/(d\chi_{0}^{-1}(T)/dT) $ vs. $ T $, respectively. The advangtage of KF plot is that no prior knowledge of $ T_{c} $ is required, which is directly defined as the intercept of such fitted straight lines on temperature axis. The KF plot for this sample is presented in Fig. \ref{Fig6}(c). From the fitted straight lines, the estimated exponents and $ T_{c} $ are $ \beta = 0.332 \pm 0.002  $, $ T_{c} = 300.21 \pm 0.05 $ K and $ \gamma = 1.18 \pm 0.01 $, $T_{c} = 300.03 \pm 0.08  $ K, which is consistent with the values from modified Arrott plot as shown in Table \uppercase\expandafter{\romannumeral1}. This implys the estimated values are self-consistent and reliable.

The critical exponents $ \beta $ and $ \gamma $ are given by the iterative modified Arrott plot, while  $ \delta $ can be obtained according to the Eq.(\ref{equ4}). Fig. \ref{Fig6}(d) shows the isothermal magnetization $ M(H) $ at the critical temperature $ T_{c} = 300 $ K, and the corresponding lg-lg scale plot is presented in the inset of Fig. \ref{Fig6}(d)}, which gives a straight line with slope $ 1/\delta $ according to Eq.(\ref{equ4}). By this means, $ \delta = 4.61 \pm 0.03 $ is obtained. To check the reliability of such analysis, $ \delta $ is also calculated by using the Widom scaling relation:\citep{PhysicsPhysiqueFizika.2.263,10.1063/1.1696618,10.1063/1.1726135}
\begin{equation}\label{equ8}
\centering
\delta = 1 + \frac{\gamma}{\beta}
\end{equation}
which gives $ \delta = 4.55 \pm 0.009 $ and $ 4.55 \pm 0.01 $, by using the $ \beta $ and $ \gamma $ obtained from modified Arrott plot and Kouvel--Fisher plot, respectively, which are coincident with those fitted by using Eq.(\ref{equ4}). From above analysis, the critical exponents ($\beta,\gamma ,$ and $\delta $) are self-consistent and unambiguous.

It is essential to check the reliability of the obtained critical exponents and $ T_{c} $ by scaling analysis.  According to the scaling hypothesis, in the asymptotic critical region, the magnetic equation of state is written as:
\begin{equation}\label{equ9}
\centering
M(H,\varepsilon) = \varepsilon^{\beta}f_{\pm}(H/\varepsilon^{\beta+\gamma})
\end{equation}
where $ f_{\pm} $ are regular functions denoted as $ f_{+} $ for $ T > T_{c} $ and $ f_{-} $ for $ T < T_{c} $. Defining the renormalized magnetization as $ m \equiv \varepsilon^{-\beta}M(H,\varepsilon) $, and the field as $ h \equiv H \varepsilon^{-(\beta + \gamma)} $, the scaling equation Eq.(\ref{equ9}) can be expressed as
\begin{equation}\label{equ10}
\centering
m = f_{\pm}(h)
\end{equation}
The scaled $ m $ plotted as a function of scaled $ h $ will fall on two universal curves: one above $ T_{c} $ and the other below $ T_{c} $, when the value of $ \beta, \gamma, $ and $ \delta $ is chosen correctly. It is an crucial criterion for the critical regime.
\begin{table*} \label{critical}
	\renewcommand\arraystretch{1.5}
	\centering
	\caption{Values of the exponents $ \beta $, $ \gamma $, and $ \delta $ as determined from the modified Arrott plots (MAP), Kuvel--Fisher plot (KF), and the critical isotherm are listed for Mn$ _{5} $Ge$ _{3} $. The theoretically predicted values of exponents for various universality classes are given for the sake of comparison. }
	\begin{threeparttable}
	\resizebox{\textwidth}{!}{
	\begin{tabular}{ccccccccc} 
		\hline
		\hline
		& Composition & Technique & Reference  & $ T_{c} $ (K) & $ \beta $ & $\gamma$ & $ \delta $& \\
		\hline
		
		& Mn$ _{5} $Ge$ _{3} $ & MAP & This work & 300.29 $ \pm $ 0.01  &  0.336 $ \pm $ 0.001  &  1.193 $ \pm $ 0.003  &   4.55 $ \pm $ 0.009$ ^{a} $  & \\
		
		& & KF & &  300.21 $ \pm $ 0.05  &  0.332 $ \pm $ 0.002  &  1.18 $ \pm $ 0.01  &  4.55 $ \pm $ 0.01$ ^{a} $ 
		&\\
		
		& & Critical isotherm & & & & &  4.61 $ \pm  $0.03  &\\
		
		& 3D-Heisenberg & Theory & \citep{KAUL19855} & & 0.365 & 1.386 & 4.8 &\\
		
		& 3D-Ising & Theory & \citep{KAUL19855} & & 0.325 & 1.24 & 4.82 &\\
		
		& 3D-XY & Theory & \citep{KAUL19855} & & 0.345 & 1.316 & 4.81 &\\
		
		& Tricritical mean field & Theory & \citep{BANERJEE196416}  & & 0.25 & 1.0 & 5.0 &\\
		
		\hline
		\hline
	\end{tabular} }
\begin{tablenotes}   
	\footnotesize              
	\item[$ a $] Calculated from Widon scaling relation $ \delta = 1 + \gamma/\beta $      
\end{tablenotes}          
\end{threeparttable} 
\end{table*}

Based on the scaling equation Eq.(\ref{equ10}), scaled $ m $ vs. scaled $ h $ is plotted in Fig. \ref{Fig6}(e) with the corresponding lg-lg scale plot in the inset. Prominently, all the data does show two separate branches below $ T_{c} $ and above $ T_{c} $. The exponents and $ T_{c} $ are further ensured with more rigorous method by plotting $ m^{2} $ vs. $ h/m $\citep{KAUL19855}, as shown in Fig. \ref{Fig6}(f), where all data also falls on two independent branches. In addition, the interactions get properly renormalized in a critical regime following the scaling equation of state, which is expressed as another form,

\begin{equation}\label{equ11}
\centering
\frac{H}{M^{\delta}} = k \left(\frac{\varepsilon}{H^{1/\beta}}\right)
\end{equation}
where $ k(x) $ is the scaling function. According to Eq.(\ref{equ11}), all experimental curves will collapse into a single curve. The inset of Fig. \ref{Fig6}(f) shows the $ MH^{1/\delta} $ vs. $ \varepsilon H^{-1/(\beta \delta)} $ for Mn$ _{5} $Ge$ _{3} $, where the experimental data collapse into a single curve with $ T_{c} $ locating at the zero point of the horizontal axis. The reliability of the obtained critical exponents are further checked by the well-rescaled curves\citep{PhysRevB.96.144429}.

The critical exponents of Mn$ _{5} $Ge$ _{3} $, obtained from different analysis techniques and different theoretical models, are listed in Table  \uppercase\expandafter{\romannumeral1} for comparison. Previous study \citep{Taroni_2008} has shown that the critical exponent $ \beta $ for a 2D magnet should be bounded within a window $ \sim $ 0.1 $ \leq \beta \leq $ 0.25 . The value of $ \beta $ in Mn$ _{5}$Ge$ _{3}$ is apparently larger than 0.25, thus exhibiting obvious 3D critical phenomenon, namely, $ d=3 $. One can see that the critical exponent $ \gamma $ of Mn$ _{5} $Ge$ _{3} $ is close to that of 3D-Ising model, and $ \beta $ approaches to that of 3D-XY or 3D-Ising model, indicating the critical behavior of Mn$ _{5} $Ge$ _{3} $ may not belong to a single universality class. Anyhow, 3D-XY and 3D-Ising models both indicate the existence of short-range magnetic interaction in Mn$ _{5} $Ge$ _{3} $. Subsequently, it is essential to understand the nature as well as the range of interaction in this compound. It is well known that the universality class of the magnetic phase transition is dictated by $ J(r) $ in the homogeneous magnet. This kind of magnetic ordering has been treated as an attractive interaction of spins, where a renormalization group theory analysis suggests $ J(r) $ decays with distance $ r $ as \citep{PhysRevLett.29.917}: 
\begin{equation}\label{equ12}
\centering
J(r) \approx r^{-(d + \sigma)} 
\end{equation}
where $ d $ stands for the spatial dimensionality and $ \sigma $ is a positive constant. Additionally, the susceptibility exponent $ \gamma $ is predicted as
\begin{equation}\label{equ13}
\begin{aligned}
\gamma \approx &  1 + \frac{4}{d}\left(\frac{n+2}{n+8}\right)\Delta\sigma+ \frac{8(n+2)(n-4)}{d^{2}(n+8)^{2}} \\
& \times \left[1+\frac{2G(\frac{d}{2})(7n+20)}{(n-4)(n+8)}\right]\Delta\sigma^{2}
\end{aligned}
\end{equation}	
where $ \Delta \sigma = (\sigma - \frac{d}{2}) $ and $ G (\frac{d}{2}) = 3 - \frac{1}{4}\left(\frac{d}{2}\right)^{2} $, $ n $ is the spin dimensionality. As $ \sigma > 2 $, the Heisenberg model applies to the 3D isotropic magnet, where $ J(r) $ decreases faster than $ r^{-5} $. As $ \sigma \leq 3/2 $, the mean-field model is satisfied, where $ J(r) $ decreases slower than $ r^{-4.5} $. To obtain the values of $ d $, $ n $, and $ \sigma $ for Mn$ _{5} $Ge$ _{3} $, the method described in Ref.\citep{PhysRevLett.29.917} is adopted in this work, where $ \sigma $ is initially adjusted according to Eq.(\ref{equ13}) with several sets of $  \{3:n\} $ to get a proper $ \gamma $ that is close to the experimental value ($ \sim $ 1.193). The obtained $ \sigma $ is then used to calculate other critical exponents by the following equations: $ \nu = \gamma/\sigma $, $ \alpha = 2 - \nu d $, $ \beta = (2-\alpha-\gamma)/2 $, and $ \delta = 1+\gamma / \beta $. This procedure is repeated for different sets of $  \{3:n\} $. Finally, $  \{3:n\}=\{3:1\} $ and $ \sigma = 1.855 $ give the critical exponents of $ \beta=0.372 $, $ \gamma=1.206 $, and $ \delta =4.240  $, which match well  with experimental values. The value of $ \sigma = 1.855 $ locates between $ 3/2$ and $2 $, which means that the spin interactions decay with distance as $ J(r)\approx r^{-4.855} $. Thus, the interactions are extended beyond the nearest neighbors, which may be ascribed to the almost same interatomic distance for \textit{NN} Mn\uppercase\expandafter{\romannumeral1}--Mn\uppercase\expandafter{\romannumeral2}, \textit{NN} Mn\uppercase\expandafter{\romannumeral2}--Mn\uppercase\expandafter{\romannumeral2}, \textit{NNN} Mn\uppercase\expandafter{\romannumeral2}--Mn\uppercase\expandafter{\romannumeral2} ($ \bigtriangleup d<0.1 $ \AA)) and the unequal magnitude of exchange strengths in Mn--Mn interactions.

Early specific heat investigation shows that the transition temperature shifts slightly towards lower temperatures with an increasing field applied below $ 0.2 $ T, similar to an antiferromagnetic transition \citep{TOLINSKI20141}. Besides the most positive exchange constants between Mn atoms, DFT calculations with the relaxed or the rigid model for Mn$ _{5} $Ge$ _{3} $ both demonstrate that there also exist a small negative exchange constant indicating the existence of antiferromagnetic interactions\citep{10.1063/1.3134482}. As is known, difference in distances between Mn atoms lead to instability of the moment, and different set of Mn--Mn interactions with unequal magnitude of exchange strengths exist in Mn$ _{5} $Ge$ _{3} $ \citep{Forsyth_1990}. Accordingly, the non-collinearity of Mn$ _{5} $Ge$ _{3} $ may be attributed to the combined effects of Mn moment instability and the coexistence of FM and small AFM coupling. These results are consistent with our resistivity measurements and previous studies\citep{10.1063/1.1633684,10.1021/nl303645k,PhysRevB.91.214425}.
Subsequently, the existence of noncollinear spin configurations leads to a small deviation of obtained critical exponents from those for standard 3D-Ising model [see Table \uppercase\expandafter{\romannumeral1}]. Therefore, our results indicate that the spin interaction in Mn$ _{5} $Ge$ _{3} $ is of 3D-Ising type coupling.

\section{Conclusions}
In summary, we have investigated the magnetic critical behavior in vicinity of the PM to FM phase transition in the itinerant ferromagnet Mn$ _{5} $Ge$ _{3} $, with its $ T_{c} $ at about 300 K. The estimated critical exponents $ \beta, \gamma $ and $ \delta $ values from  various techniques and theoretical models show good consistency with each other and follow the scaling behavior well, confirming that the obtained exponents are unambiguous and intrinsic. The critical exponents suggest a second order phase transition and are close to the values for the 3D-Ising model. The magnetic exchange  is found to decay as $ J(r)\approx r^{-4.855} $ and the spin interactions are extended beyond the nearest neighbors, which may be related to the close interatomic distance among Mn atoms and different set of Mn-Mn interactions with unequal magnitude of exchange strengths. Furthermore, the existence of noncollinear spin configurations in Mn$ _{5} $Ge$ _{3} $ leads to a small deviation of obtained critical exponents from those for standard 3D-Ising model.

\section{Acknownledgments}
This work is supported by the National Natural Science Foundation of China (Grant Nos. 11874422 and 12074425), the National Key R\&D Program of China (Grant No. 2019YFA0308602), and the Research Funds of Renmin University of China (Grant No. 19XNLG18).

\bibliography{Bibtex}
\end{document}